\documentclass[aps,prl,twocolumn,superscriptaddress]{revtex4-1}
\usepackage{amsmath}
\usepackage{physics}
\usepackage{graphicx}
\DeclareMathOperator{\sign}{sign}

\newcommand{\be}{\begin{equation}}
\newcommand{\ee}{\end{equation}}
\newcommand{\bea}{\begin{eqnarray}}
\newcommand{\eea}{\end{eqnarray}}

\def\up{\uparrow}
\def\dn{\downarrow}

\providecommand{\up}{\uparrow}
\providecommand{\dn}{\downarrow}

\newcommand{\para}[1]{\left(#1\right)}

\newcommand{\COMMENT}[1]{}

\begin{document}

\title{Enhanced correlations and superconductivity in weakly interacting partially flat band systems: a determinantal quantum Monte Carlo study}
\author{Edwin W. Huang}
\affiliation{Department of Physics, Stanford University, Stanford, CA 94305, USA}
\affiliation{Stanford Institute for Materials and Energy Sciences, SLAC National Accelerator Laboratory and Stanford University, Menlo Park, CA 94025, USA}
\author{Mohammad-Sadegh Vaezi}
\affiliation{Pasargad Institute for Advanced Innovative Solutions (PIAIS) , Tehran 1991633361, Iran}
\affiliation{Department of Physics, Washington University, St. Louis, MO 63160, USA}
\author{Zohar Nussinov}
\affiliation{Department of Physics, Washington University, St. Louis, MO 63160, USA}
\author{Abolhassan Vaezi$^*$}
\affiliation{Department of Physics, Stanford University, Stanford, CA 94305, USA}
\affiliation{Stanford Center for Topological Quantum Physics, Stanford University, Stanford, CA, USA}
\email{vaezi@stanford.edu}

\date{\today}

\begin{abstract}
Motivated by recent experiments realizing correlated phenomena and superconductivity in 2D van der Waals devices, we consider the general problem of whether correlation effects may be enhanced by modifying band structure while keeping a fixed weak interaction strength. Using determinantal quantum Monte Carlo, we study the 2D Hubbard model for two different band structures: a regular nearest-neighbor tight-binding model, and a partially flat band structure containing a non-dispersing region, with identical total non-interacting bandwidth $W$. For both repulsive and attractive weak interactions ($\abs{U} \ll W$), correlated phenomena are significantly stronger in the partially flat model. In the repulsive case, even with $U$ an order of magnitude smaller than $W$, we find the presence of a Mott insulating state near half-filling of the flat region in momentum space. In the attractive case, where generically the ground state is superconducting, the partially flat model exhibits significantly enhanced superconducting transition temperatures. These results suggest the possibility of engineering correlation effects in materials by tuning the non-interacting electronic dispersion.  
\end{abstract}

\maketitle

\begin{figure}
\centering
\includegraphics[height = 3.5cm, width=8.5cm]{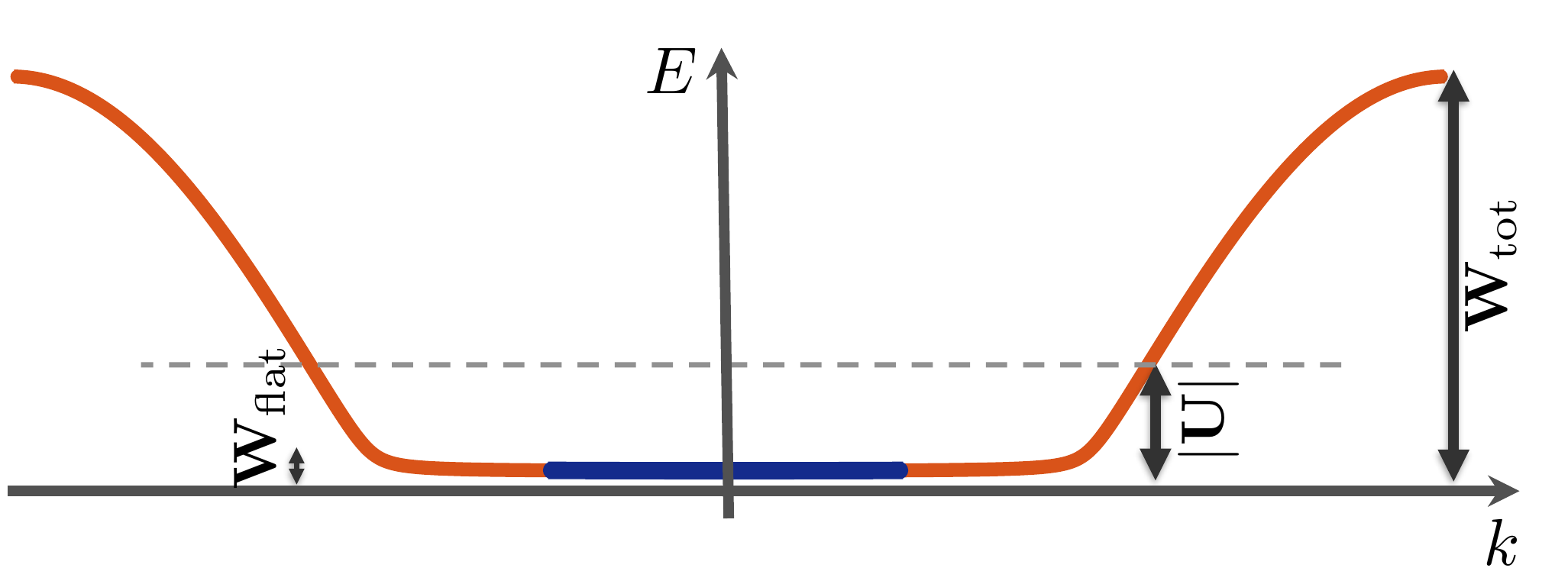}
\caption{A schematic band structure of partially flat band systems. The band structure contains a (nearly) flat region with a high DOS and a narrow bandwidth $W_{\rm flat}$. In the PFB system, the interaction energy scale may be much smaller (larger) than the total (flat region) bandwidth; thus: $W_{\rm flat} \ll \abs{U} \ll W_{\rm tot}$. The blue region denotes occupied energy states when the interaction is switched off. However, using DQMC we find that all single particle states inside the flat region are (almost) equally occupied upon considering interaction effects (see the text). Due to the absence of any mass gap between the flat and dispersive areas, the two regions are strongly coupled through interactions.}
\label{fig:0}
\end{figure}

\noindent {\bf Introduction.--} The recent discovery of superconductivity in the twisted bilayer graphene (TBG)~\cite{Cao2018_1,Cao2018_2} has spurred increasing interest in 2D van der Waals materials with structural deformations~\cite{novoselov2005two,zhang2005experimental,guinea2010energy,ju2015topological,vaezi2013topological_a,zhang2013valley,martin2008topological,wang2012electronics} and has inspired new venues to search for high $T_c$ superconductivity~\cite{bednorz1986possible,wu1987superconductivity,anderson1987resonating,fradkin2015colloquium,lee2006doping,kuroki2008unconventional,mazin2008unconventional,kamihara2008iron}. In TBG, the band structure hosts tiny regions near $K$ and $K'$ valleys with nearly flat energy dispersions~\cite{morell2010flat,bistritzer2011moire,mele2010commensuration,trambly2010localization,fang2016electronic,zhang2018low}. When these regions are partially occupied, a phase diagram similar to that of high $T_c$ cuprates has been reported \cite{Cao2018_1,Cao2018_2}. Alongside and possibly compounding other effects, e.g., \cite{Phil}, it is widely believed that due to the large density of states (DOS) at the two tiny (nearly) flat regions, the system exhibits strong correlation physics~\cite{Cao2018_2,heikkila2016flat,volovik2018graphite,wu2018hubbard,yuan2018model,po2018origin,dodaro2018phases,roy2018unconventional,rademaker2018charge,xu2018kekul,xu2018topological,guo2018pairing}. Inspired by these experiments on TBG and the broader quest of understanding flat bands, we introduce ``Partially Flat Band" (PFB) models wherein the band structure is neither fully flat nor fully dispersive (see Fig. \ref{fig:0}). In PFBs, the bare kinetic (i.e., not interaction induced \cite{Khodel,Review_condensate}) dispersion $\epsilon_{\mathbf k}$ is nearly flat over a finite fraction of the Brillouin zone (BZ) with a diverging DOS.

There are no reliable theoretical tools to obtain the effective low energy action for the PFB. Perturbation theory fails due to the divergence of the DOS over a finite portion of the BZ. Other conventional methods of strongly correlated systems such as the Wolff-Schrieffer transformation become inapplicable. These difficulties are tied to the existence of three significantly different energy scales:  (i) the bandwidth associated with the flat region, (ii) the total bandwidth, and (iii) the interaction energy scale which is much greater than (i) yet far smaller than (ii). Interactions can mix the smoothly connected flat and dispersive regions. These two regions may actively exchange particle and energy. Thus, if the nearly flat region is partially filled, the dispersive region cannot be disregarded. A projection of the interactions onto the flat region is unjustified.

In order to better grasp the physics of PFB systems, we introduce a toy model allowing numerical studies on general lattices. Herein, a large fraction (of order one) of the band structure is (nearly) flat.
We utilize the Determinental Quantum Monte Carlo (DQMC) approach to obtain the phase diagram for both repulsive and attractive weak Hubbard interactions. Due to the existence of flat areas, correlation effects are pronounced and we expect to encounter evidence of strong correlation physics despite only weak interactions. In particular, we observe an emergent Mott insulating state near {\bf half-filling} of the {\bf flat region} in momentum space. Our calculations show that the momentum space electron occupation number becomes nearly uniform and fractional all over the flat region. This is inconsistent with the Luttinger theorem and constitutes another indication that we either have a gapless non-Fermi liquid or a Mott insulating phase.
Lastly, we find a considerable enhancement of the superconducting transition temperature for attractive interactions which can be generated via, e.g., retarded phonon-mediated electron-electron coupling~\cite{dodaro2018phases,peltonen2018mean,wu2018theory,lian2018twisted}.

\noindent {\bf Model.--} The Hubbard model Hamiltonian is given by 
\begin{equation}
H = \sum_{\mathbf k\sigma} \epsilon_{\mathbf k} c_{\mathbf k \sigma}^\dagger c_{\mathbf k \sigma} + U \sum_{i} n_{i \uparrow} n_{i \downarrow}.\label{eq1}
\end{equation}
Here, $c_{\mathbf k \sigma}^\dagger$ creates an electron of momentum $\mathbf k$ and spin $\sigma$, the (non-interacting) band dispersion is $\epsilon_{\mathbf k}$, and $n_{i \sigma} = c_{i \sigma}^\dagger c_{i \sigma}$ is the number operator on site $i$. The local (on-site) interaction is parameterized by $U$. Thanks to its possible relevance to high-$T_c$ cuprate superconductors, the repulsive ($U>0$) Hubbard model on a square lattice has been the focus of  many numerical studies. Due to the fermion sign problem in quantum Monte Carlo (QMC) simulations of the repulsive Hubbard model, unbiased numerical results are absent at temperatures relevant to the putative superconducting phase of the model (though a variety of techniques suggest the presence of $d$-wave superconductivity and various competing phases). By contrast, the attractive Hubbard model ($U < 0$) is amenable to sign-problem-free QMC simulations, allowing for detailed characterization of the $s$-wave superconducting phase, including calculation of $T_c$. We will study both the repulsive and attractive realizations of this model. 

For simplicity and to ease comparison to existing studies of Hubbard models, we performed the simulations on the commonly studied periodic $L \times L$ square lattice geometries. Here, the band structure 
\begin{align}
\epsilon_{\mathbf k} &= (1 + f\sign(\epsilon_{\mathbf k}^0)) \epsilon_{\mathbf k}^0 \\
\epsilon_{\mathbf k}^0 &= -2(\cos k_x + \cos k_y). \label{eq3}
\end{align}
The parameter $f$ controls the flatness of the band. For $f = 0$, $\epsilon_{\mathbf k} = \epsilon_{\mathbf k}^0$ corresponding to nearest neighbor hopping dispersion. Here,  both the nearest neighbor hopping amplitude $t$ and the lattice constant are set to unity. (In this work, all energies will be given in units of $t(=1)$ and we will further set the Boltzmann constant $k_{B}$ to unity.) For $f=1$, the dispersion $\epsilon_{\mathbf k} = 0$ when $\epsilon_{\mathbf k}^0 \leq 0$ (half the BZ) and $\epsilon_{\mathbf k} = 2 \epsilon_{\mathbf k}^0$ otherwise. We refer to the $f=0$ model as the ``regular band'' Hubbard model and to the $f=1$ system as the ``PFB Hubbard model''. 

Importantly, the total bandwidth is fixed to $W_{tot} = 8$ in either case. Hence, our data showcases the effects of introducing a flat region in the non-interacting dispersion, while keeping the total bandwidth constant. In all cases, we will focus on the hole-doped model (i.e., an  average occupancy $\langle n \rangle = \langle n_\uparrow + n_\downarrow \rangle < 1$), such that if $f=1$, the non-interacting Fermi level lies inside the flat region. 

\begin{figure}
\centering
\includegraphics{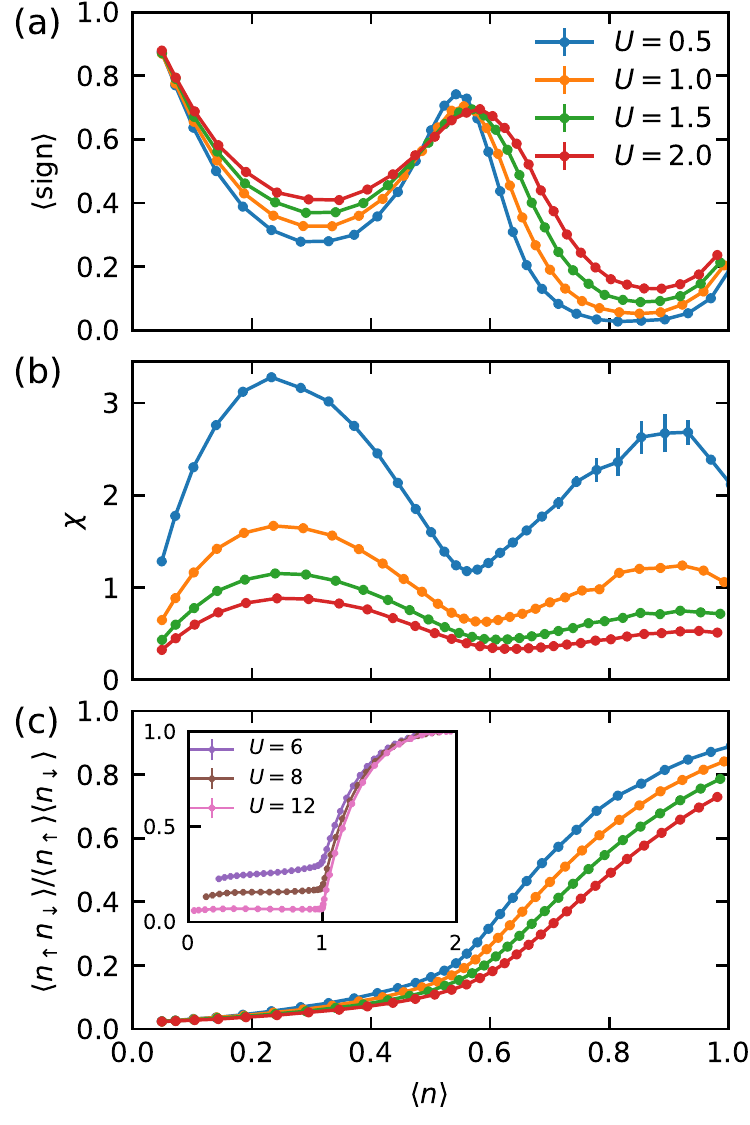}
\caption{Doping dependence of average fermion sign (a), charge compressibility $\chi = \frac{\partial \langle n \rangle}{\partial \mu}$ (b), and double occupancy ratio (c) in the partially flat band model with a repulsive interaction $U > 0$. In the inset of (c), we plot the double occupancy ratio for the regular band Hubbard model with strong interactions. All simulations here are for a $16 \times 8$ periodic cluster at temperature $T = U/15$. All error bars are $\pm 1$ standard error of the mean, determined by jackknife resampling.}
\label{fig:1}
\end{figure}

\noindent {\bf Repulsive interaction.--} We first consider the repulsive model with a partially flat band and weak interactions $U \leq 2$. The presence of the fermion sign problem restricts accessible temperatures to $T \gtrsim U/15$ for moderate system sizes ($\sim 100$ sites), with certain fillings amenable to somewhat lower temperatures. Interestingly, the average sign in the DQMC simulation is enhanced near a density of $\langle n \rangle \sim 0.6$ per unit cell and decreases rapidly away from this value [Fig.~\ref{fig:1}(a)]. This behavior is reminiscent of that in the repulsive Hubbard model with a regular band, where the sign is protected by particle-hole symmetry at exactly half-filling ($\langle n \rangle = 1$); doping away from half-filling (such that $\langle n \rangle \neq 1$) leads to a severe sign problem. While no such symmetry is exactly manifest in the partially flat model, the similar behavior of the $\langle n \rangle \sim 0.6$ PFB system to the regular Hubbard model at half-filling hints at similar (Mott insulating) underlying physics.

\begin{figure}
\centering
\includegraphics{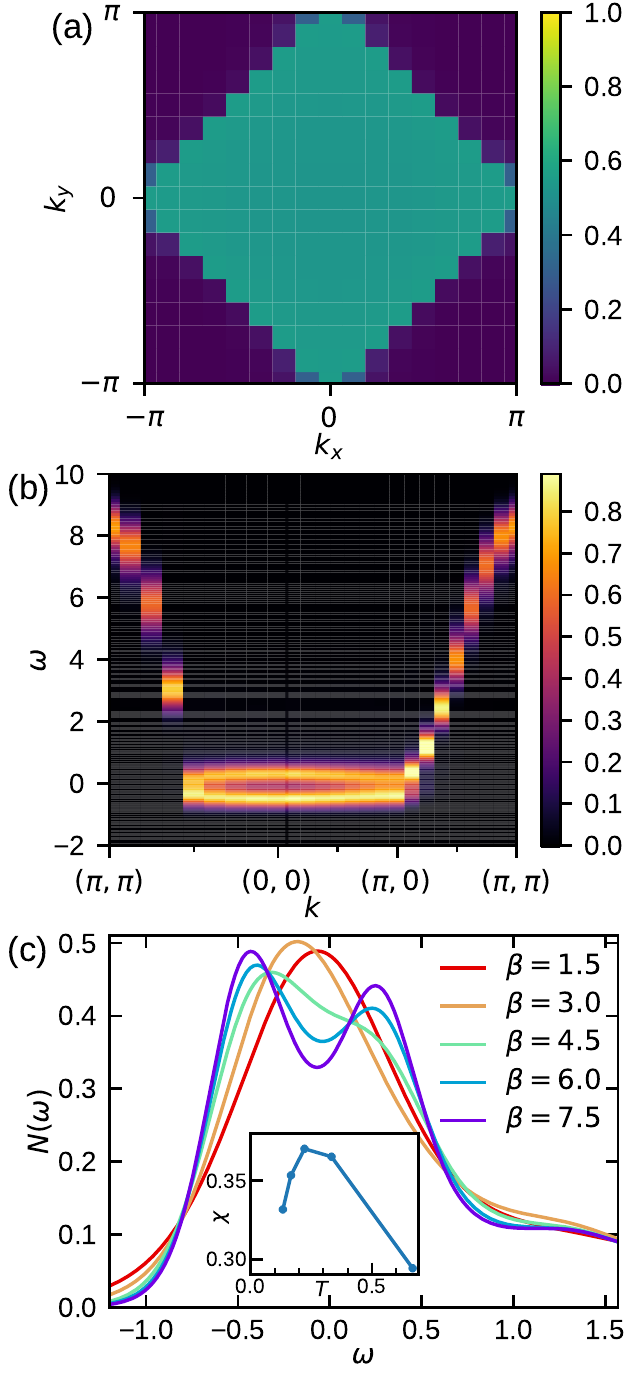}
\caption{(a) Momentum resolved electron filling $\langle n_{\mathbf k} \rangle = \langle c_{\mathbf k}^\dagger c_{\mathbf k} \rangle$ for the partially flat band model with repulsive interaction $U = 2$ at temperature $T = 0.133$ and average filling $\langle n \rangle = 0.62$, on a periodic $16 \times 16$ cluster. (b) Single particle spectral function $A({\mathbf k},\omega)$ along high-symmetry cuts obtained for the same simulation using maximum entropy analytic continuation. (c) Single particle density of states $N(\omega)$ for the same parameters at inverse temperatures $\beta=1/T$ as given in the legend. Inset: charge compressibility as a function of temperature.}
\label{fig:2}
\end{figure}

To confirm this, we examine the charge compressibility $\chi = \frac{\partial \langle n \rangle}{\partial \mu}$ in Fig.~\ref{fig:1}(b). As a function of filling, $\chi$ has a pronounced dip around $\langle n \rangle \sim 0.6$. The compressibility at $\langle n \rangle = 0.62$ decreases with lowering temperature [Fig.~\ref{fig:2}(c), inset], indicating insulating behavior and suggesting an incompressible gapped ground state. To relate this behavior to Mott physics and quantitatively assess correlation effects, we compare the number of doubly occupied sites $\langle n_\uparrow n_\downarrow \rangle$ to the uncorrelated case (in which $\langle n_\uparrow n_\downarrow \rangle = \langle n_\uparrow \rangle \langle n_\downarrow \rangle = \langle n \rangle^2/4$). The ensuing ratio is plotted in Fig.~\ref{fig:1}(c). For a regular Hubbard model [inset of Fig.~\ref{fig:1}(c)], at half-filling, this ratio is suppressed when there are strong interactions. This ratio remains suppressed upon hole-doping but rises for electron-doping where double occupancy becomes unavoidable. In the PFB model, we observe the same behavior relative to a filling of $\langle n \rangle \sim 0.6$ at which a crossover occurs.  
The suppression, even for $U$  $\sim$ $1$, of double occupancy in the PFB system is comparable in magnitude to that of the regular band Hubbard model with $U \sim 8$. Taken together, the analogies between the weakly interacting PFM model and the regular band strongly interacting Hubbard model demonstrate that even weak interactions are sufficient to enable correlated phenomena given the correct band structure.

Knowing that a Mott insulating state appears in the PFB model when $\langle n \rangle \sim 0.6$, we now explore in greater depth the momentum and energy dependence of the single-particle properties. In Fig.~\ref{fig:2}(a), we plot the electron occupancy $\langle n_{\mathbf k} \rangle = \langle c_{\mathbf k}^\dagger c_{\mathbf k} \rangle$. As is evident from Eq.~\ref{eq3}, the non-dispersive flat region is delineated by $\abs{k_x} + \abs{k_y} \leq \pi$. In this region, the electron occupancy varies from $0.52$ to $0.55$ while the total filling (Fig.~\ref{fig:2}(a)) is $\langle n \rangle = 0.62$. Thus, the crossover seen in Fig.~\ref{fig:1} at $\langle n \rangle \sim 0.6$ corresponds to a half-filling of the flat portion of the PFB. Inconsistency with Luttinger's theorem implies a non-Fermi liquid type behavior or a featureless gapped (Mott insulating) state. This special behavior of the occupancy suggests a modified mean-field approach to the PFB for the effective low energy action from which the emergence of the Mott insulator becomes obvious \cite{SM}.

To corroborate these statements, we computed the single-particle spectral function $A(\mathbf k, \omega)$ by an analytical continuation of the imaginary time Green's function using the Maximum Entropy method \cite{Jarrel_analytic_QMC_1996a}. Fig.~\ref{fig:2}(b) shows $A(\mathbf k, \omega)$ along high-symmetry cuts of the BZ. The most pertinent feature is the presence of a Mott gap throughout the flat region. In Fig.~\ref{fig:2}(b), for $U=2$ and a temperature $T=0.133$,  the gap is largest ($\sim 0.8$) at $\mathbf k = (0,0)$, and gradually drops near the boundaries of the flat region. Fig.~\ref{fig:2}(c) provides the single-particle density of states $N(\omega) = \frac{1}{L^2} \sum_{\mathbf k} A(\mathbf k, \omega)$ for different temperatures; the gap opening temperature is estimated to be between $T=0.22$ and $0.33$ concomitant with the onset of insulating behavior in the charge compressibility [inset].

\begin{figure}
\centering
\includegraphics{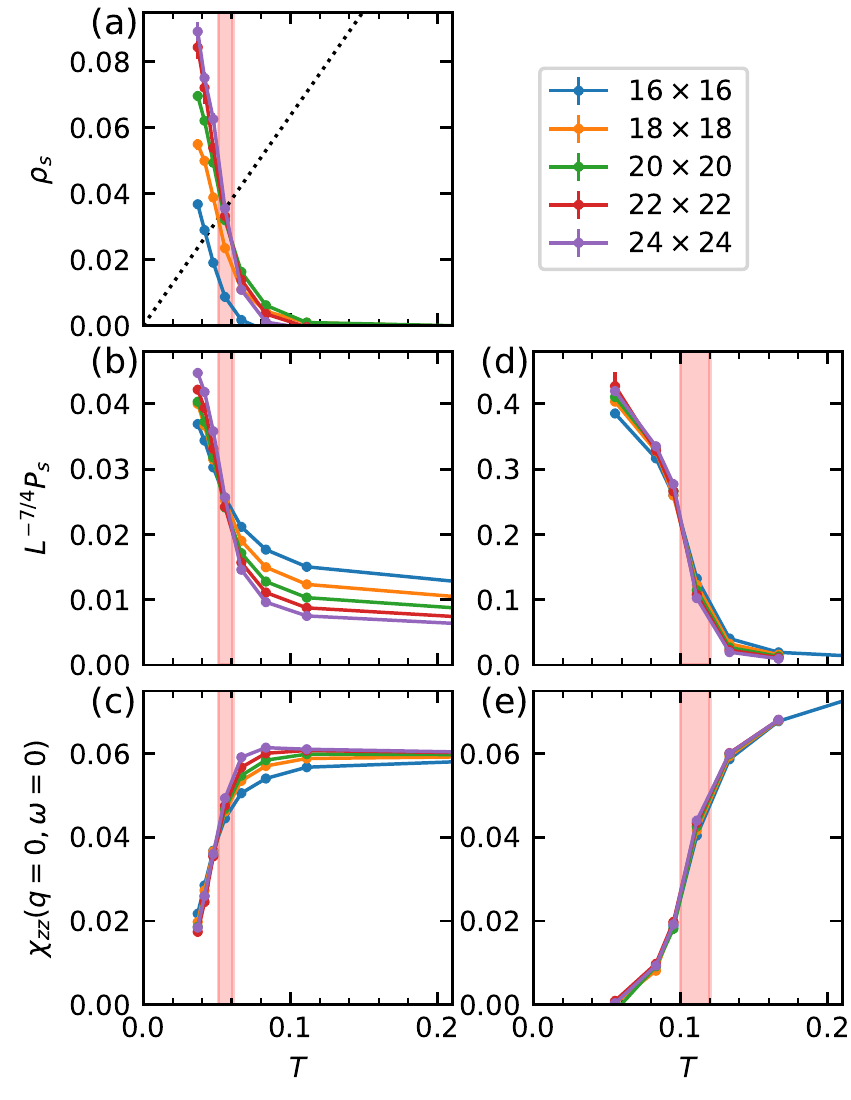}
\caption{Estimates of the superconducting transition temperature $T_c$ for the regular attractive (a,b,c) and PFB (d,e) Hubbard models. Here, $U=-2$ and $\langle n \rangle = 0.8$. We plot the superfluid stiffness $\rho_s$ (Eq. (\ref{rho_s})) in (a), the $s$-wave pair-field susceptibility (Eq. (\ref{PS})) multiplied by $L^{-7/4}$ (b,d), and the static spin susceptibility (Eq. (\ref{chi_zz})) (c,e). The dashed line in (a) is $2T/\pi$. The shaded regions indicate estimates of $T_c$.}
\label{fig:3}
\end{figure}

\begin{figure}
\centering
\includegraphics{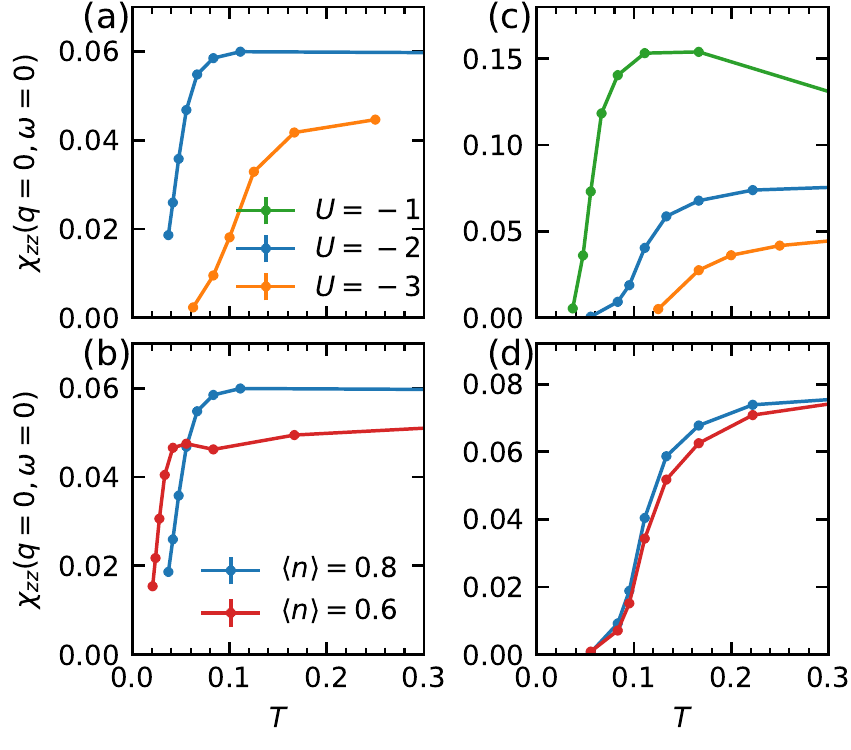}
\caption{Parameter dependence of static spin susceptibility in the regular attractive Hubbard model (a,b) and in the partially flat band model (c,d). The downturn in the static spin susceptibility signals formation of singlet pairs and provides a rough indication of $T_c$.}
\label{fig:4}
\end{figure}

\noindent {\bf Attractive interaction.--} While we have found strong indications of Mott insulating physics in the repulsive partially filled model with only weak interactions, the sign problem prevents a detailed study of phases that emerge by doping away from $\langle n \rangle \sim 0.6$. By contrast, by studying the attractive Hubbard model via DQMC \cite{BSS}, we can establish concrete results on the effect of the modified band structure on superconductivity. In 2D simulations, accurate estimates of superconducting $T_c$ may be obtained using the Nelson-Kosterlitz (NK) criterion for superfluid stiffness: $\rho_s(T_c) = 2 T_c/\pi$. For finite cluster simulations, the temperature where $\rho_s(T)$ intersects with $2 T/\pi$ provides an estimate for $T_c$ in the thermodynamic limit. In DQMC, $\rho_s$ may be calculated as \cite{Paiva}
\begin{align}
\label{rho_s}
\rho_s &= \frac{1}{4}\left(\Lambda^L - \Lambda^T\right) \\
\Lambda^L &= \lim_{q_x \to 0} \Lambda_{x x}(q_x, q_y=0, \omega_n=0) \\
\Lambda^T &= \lim_{q_y \to 0} \Lambda_{x x}(q_x=0, q_y, \omega_n=0),
\end{align}
where
\begin{align}
\Lambda_{x x}(\mathbf q, \omega_n) = \sum_{i} \int_0^\beta d\tau\, e^{-i(\mathbf q \cdot \mathbf r_i - \omega_n \tau)} \langle j_{x}(i, \tau) j_{x}(0, 0) \rangle.
\end{align}
Here, $j_{x}(i)$ is the $x$ component of the current density operator at lattice site $i$, $\mathbf j(i) = \sum_{j \sigma} i t_{i j} \left( \mathbf r_i - \mathbf r_j \right) c_{i \sigma}^{\dagger} c_{j \sigma}$ where $t_{i j}$ is the hopping between sites $i$ and $j$ (related to the dispersion through $t_{i j} = -\frac{1}{L^2} \sum_{\mathbf k} e^{i \mathbf k \cdot (\mathbf r_i - \mathbf r_j)} \epsilon_{\mathbf k}$.)

We show the results of this analysis in Fig.~\ref{fig:3}(a), for the attractive regular band Hubbard model for $U=-2$, and $\langle n \rangle = 0.8$. Comparing simulations on different cluster sizes allows us to estimate $T_c \approx 0.056(5)$ for these parameters. Here, the minimal cluster size for a reasonable estimate of $T_c$ is $\sim 20 \times 20$. (In a previous DQMC simulation of the attractive Hubbard model \cite{Paiva} for $U = -4$, a lattice of size $\sim 10 \times 10$ was sufficient for estimating $T_c$. For low $|U|$, the longer superconducting coherence length requires larger clusters to mitigate finite-size effects.)

A PFB requires many real-space hopping amplitudes to be non-zero. Consequently, the computation of the current correlator becomes expensive. As an alternative, we infer $T_c$ from the behavior of the pair field susceptibility and of the static spin susceptibility. The equal-time $s$-wave pair field susceptibility is given by
\begin{equation}
\label{PS}
P_s = \langle \{\Delta, \Delta^\dagger\} \rangle,
\end{equation}
where $\Delta^\dagger = \frac{1}{L} \sum_{i} c_{i \uparrow}^\dagger c_{i \downarrow}^\dagger = \frac{1}{L} \sum_{\mathbf k} c_{\mathbf k \uparrow}^\dagger c_{-\mathbf k \downarrow}^\dagger$ is the $s$-wave pair field creation operator at zero net momentum.
The spin susceptibility is given by
\begin{equation}
\label{chi_zz}
\chi_{z z}(\mathbf{q}, \tau) = \langle T_\tau S_z(\mathbf{q}, \tau) S_z^\dagger(\mathbf{q}) \rangle,
\end{equation}
where $S_z(\mathbf q) = \frac{1}{L} \sum_{i} e^{-i \mathbf{q} \cdot \mathbf r_i} \left(c_{i \uparrow}^\dagger c_{i \uparrow} - c_{i \downarrow}^\dagger c_{i \downarrow}\right)$. We focus on the static spin susceptibility at $\mathbf q = 0$: $\chi_{z z}(\mathbf q, \omega=0) = \int_0^\beta d\tau\, \chi_{z z}(\mathbf q, \tau)$. (We consider only the $z$ component of spin; $\chi_{x x}$, $\chi_{y y}$, and $\chi_{z z}$ are identical within statistical errors).

Upon cooling below $T_c$, one expects that the formation of singlet pairs suppresses the static spin susceptibility. In the absence of a pseudogap, the onset temperature of this suppression provides an estimate of $T_c$. A corresponding rise of the pair field susceptibility would confirm that the suppression of spin susceptibility is due to the onset of superconductivity.

Figs.~\ref{fig:3}(a-e) display the results of DQMC calculations for the temperature dependence of the pair field susceptibility and the static spin susceptibility. For the regular band, we observe the expected downturn in spin susceptibility and rise in pair field susceptibility near $T_c \approx 0.056(5)$. Similar behavior occurs in the PFB model at $T_c \approx 0.11(1)$, indicating that the attractive Hubbard model with a PFB has doubled the superconducting transition temperature of the model with a regular dispersion. This increase is partially anticipated by the larger density of states in the PFB. We emphasize that this is while keeping the interaction strength and the total noninteracting bandwidth fixed.

An enhancement of superconductivity in the attractive PFB appears for different interaction strengths and dopings. In Figs.~\ref{fig:4}(a,b), we vary the interaction strength for both the regular band and the PFB models. As before, the downturn in the static spin susceptibility provides a rough indication of $T_c$.  When $|U|=3$, the Hubbard model $T_c$ rises to $0.12(2)$ for the regular band model and $0.18(2)$ for the FBM. For a smaller interaction strength of $|U|=1$, the PFB model has a $T_c$ of $0.06(5)$ while the $T_c$ of the regular band model was too low to be readily accessible. In Figs.~\ref{fig:4}(c,d), we contrast the effects of additional hole doping within the two models. As the number density varies from $\langle n \rangle = 0.8$ to $\langle n \rangle = 0.6$,  the regular band Hubbard model $T_c$ decreases from $0.056(5)$ to $0.035(5)$ while the PFB model shows little variation its $T_c$.

\noindent {\bf Conclusions.--}
We introduced and studied PFB systems. PFBs may be realized in diverse experimental arenas, e.g., TBG or heavy fermion systems. Our DQMC analysis illustrates that the existence of flat subregions enhances the correlation effects even for interactions significantly weaker than the total bandwidth. We found a Mott insulating state for weak local repulsion and an $s$-wave superconductor with a considerably enhanced $T_c$ for weak local attraction. Our PFB model may aid the understanding of TBG whose band structure hosts extremely tiny (nearly) flat areas due to the very large spatial extent of the Moir{\'e} super-lattices. Studying systems with such super-cells is not computationally feasible. As we discussed earlier, the dispersive non-flat regions that are connected to small flat domains of the TBG may not be ignored. Thus, a projection of the Hamiltonian onto the flat region is not possible. One needs to keep single particle (hole) excitations with energies of order the interaction scale $\abs{U}$ above (below) the flat region. Our PFB model captures these essential features and provides a simple toy model to study TBG which is computationally feasible as well (albeit by imposing triangular lattice symmetry). 

The ideal PFB (i.e., the model exhibiting exactly flat subregions of the band) requires the existence of finite hopping amplitudes between distant sites. Nonetheless, we may truncate these amplitudes beyond a cutoff distance without impacting the low energy physics. In doing so, we may still achieve nearly flat regions with enhanced correlation. Remarkably, augmenting a nearest neighbor hopping tight binding amplitude ($t=1$) by an additional next nearest neighbor hopping amplitude $t_2\approx -0.54$ suffices to achieve a high DOS in the lower half of the band structure on the square lattice. Such a simple model might be realizable in 2D van der Waals devices with square lattice symmetry or in cold atoms systems via photo-induced coupling experiments or through applying pressure, and is expected to have an amplified superconducting transition temperature.

\noindent {\bf Acknowledgements.--}
We acknowledge helpful discussions with Hideo Aoki, Sharareh Sayyad and Hongchen Jiang. EWH was supported by the U.S.~Department of Energy (DOE), Office of Basic Energy Sciences, Division of Materials Sciences and Engineering, under Contract No.~DE-AC02-76SF00515. Computational work was performed on the Sherlock cluster at Stanford University. MSV and ZN acknowledge partial support by the National Science Foundation (NSF 1411229). MSV also acknowledges the financial support from Pasargad Institute for Advanced Innovative Solutions (PIAIS) under Supporting Grant scheme (Project SG1-RCM1810-01).

%

\newpage
\pagebreak

\newpage
\onecolumngrid
\pagebreak

\newpage

\renewcommand{\thefigure}{S\arabic{figure}}
\renewcommand{\theequation}{S\arabic{equation}} 
\setcounter{figure}{0}
\setcounter{equation}{0}
\setcounter{page}{1}

\begin{center}
{\large \bf Supplemental Material for ``Enhanced correlations and superconductivity in weakly interacting partially flat band systems: a determinantal quantum Monte Carlo study''}\\
\vspace{0.4cm}

Edwin W. Huang,$^{1,2}$ Mohammad-Sadegh Vaezi, $^{3,4}$ , Zohar Nussinov, $^{4}$,  and Abolhassan Vaezi$^{1,5,*}$

{\small \it $^1$ Department of Physics, Sanford University, Stanford, CA 94305, USA}

{\small \it $^2$  Stanford Institute for Materials and Energy Sciences, SLAC National Accelerator Laboratory and Stanford University, Menlo Park, CA 94025, USA}

{\small \it $3$ Pasargad Institute for Advanced Innovative Solutions, Tehran, Iran}

{\small \it $4$ Department of Physics, Washington University, St. Louis, MO 63160, USA}

{\small \it $5$  Stanford Center for Topological Quantum Physics, Stanford University, Stanford, CA, USA}

\end{center}
\vspace{0.4cm}

\subsection{A modified mean-field approach to the partially flat band systems}
We now discuss a modified mean field theory that can successfully explain quintessential features of PFB systems, e.g., the emergence of the Mott insulator near the half-filling of the flat region. The essential ingredient is the fact that the occupation number of the single particle energy eigenstates does not follow the Fermi-Dirac distribution since we have a Fermi volume rather than a Fermi surface. Instead, the $k-$ space occupancy is uniform over the flat region (where the chemical potential crosses).Thus, all associated flat band states are partially occupied. 

Motivated by the physics of square lattice regular Hubbard model near half-filling, we focus on the anti-ferromagnetic order. We assume that $\left<\frac{n_{i,\up}-n_{i,\dn}}{2}\right>= (-1)^i m$, where $m$ denotes the staggered magnetization. We invoke the standard mean field approximation $n_{i,\up}n_{i,\dn} \approx \left<n_{i,\up}\right>n_{i,\dn} + n_{i,\up}\left<n_{i,\dn}\right> - \left<n_{i,\up}\right>\left<n_{i,\dn}\right>$. Plugging this approximation into the model Hamiltonian, Eq. \ref{eq1} of the main text, and performing a Fourier transformation, we obtain

\bea
H_{\rm MF} = &&\sum_{k,\sigma} (\para{\epsilon_{k}-\mu} c_{k,\sigma}^\dag c_{k,\sigma} - mU \sigma c_{k+Q,\sigma}^\dag c_{k,\sigma} + h.c.),
\eea
where $Q = \para{\pi,\pi}$. The above mean-field Hamiltonian can be readily diagonalized. We then have

\bea
H_{\rm MF} = &&\sum_{\abs{k_x}+\abs{k_y}\leq \pi,\sigma} (E_{+,k} \gamma_{+,k}^\dag \gamma_{+,k} + E_{-,k}\gamma_{-,k,\sigma}^\dag \gamma_{-,k,\sigma}),
\eea
where $E_{\tau,k}=\tau \sqrt{\epsilon_k^2+\para{Um}^2}-\mu$ denotes the energy eigenvalue associated with band $\tau=\pm$ at momentum $k$ and $\gamma_{\tau,k}$ the corresponding annihilation operator which is a linear combination of $c_{k,\sigma}$ and $c_{k+Q,\sigma}$. The self-consistency of our assumption about the staggered magnetization implies the following identity:

\bea
m = \sum_{i}\frac{(-1)^i}{2N_s} \left<n_{i,\up}-n_{i,\dn}\right>_{H_{\rm MF}}  = -mU \sum_{k,\tau} \frac{f\para{E_{\tau,k}}}{E_{\tau,k}}.~~~
\eea
Here, $f\para{E_{\tau,k}}$ denotes the occupation number of energy band $\tau$ at momentum $k$, and $N_s =L^2$ is the number of sites. Normally, $f$ is replaced by the Fermi-Dirac distribution so that all negative energy states (those below the chemical potential) are fully occupied at $T=0$, and excited states (those above the chemical potential) are empty. In PFBs, the chemical potential crosses many zero-energy states (more than the total electron density), and thus it is not, a priori, clear which states are occupied or empty. This feature can generally lead to exotic behaviors in flat band systems such as the fractional quantum Hall systems. However, our DQMC study of the PFB system shows that the occupation number is nearly uniform all over the flat sub-region where the chemical potential is tuned. We have verified that this remains the case even for nearly flat sub-regions (so long as the interaction scale is larger than the bandwidth $W_{flat}$ of the sub-region with a high DOS, see Fig. \ref{fig:0} of the main text). Implementing this observation into the $f\para{E_{\tau,k}}$ functional, we observe that, different from the regular band Hubbard model, the mean-field anti-ferromagnetic order parameter remains finite even away from the half-filling (with the filling fraction being relative to that of the total band). 

In the conventional band Hubbard model, the Fermi-Dirac distribution can rationalize the appearance of anti-ferromagnetic order at half-filling. However, at the mean field level, any doping away from half-filling (relative to the entire band structure), even if infinitesimal, will eradicate the antiferromagnetic order. Using the modified mean-field approximation, we find that although the flat subregion is partially occupied, the staggered magnetization is non-zero. Consequently, the anti-ferromagnetic spin-density wave (SDW) will open up a mass gap separating the (nearly) flat sub-region from the dispersive sub-regions. In other words, an interaction-induced SDW mass gap will appear at the $\abs{k_x}+\abs{k_y} \leq \pi$ surface corresponding to half-filling. Thanks to the existence of finite gap separating the modified (nearly) flat region from the remaining band structure, we can focus on the lower flat sub-band and employ the standard techniques of the strongly correlated system on a modified (nearly) flat sub-band. One consequence of this simple analysis is that the system will exhibit a Mott insulating phase at half-filling of the lower (nearly) flat emergent sub-band (i.e., at a quarter filling of the original band). 

To summarize, the interaction generates an SDW order, doubles the unit cell, and the relevant flat sub-band around the chemical potential becomes separated from other bands (which were otherwise smoothly connected to the flat sub-band in the absence of interaction). Note that the bandwidth and structure factors of the emergent (nearly) flat sub-band differ from those of the original flat subregion. The emergent well-separated (nearly) flat sub-band is not fully occupied and the interaction projected the emergent flat sub-band will dictate its fate. Mott physics, as well as other related strong correlation phenomena, are possible. This picture can be easily generalized to more complicated situations where the (nearly) flat subregion is smaller by considering smaller nesting vectors (different and shorter $Q$ vectors) that lead to larger unit cells. 


\end{document}